\documentclass[11pt]{article}

\usepackage[english,francais]{babel}

\usepackage{graphicx}

\newfont{\ensmathquatorze}{msbm10 scaled 1400}
\newfont{\ensmathonze}{msbm10 scaled 1100}
\newfont{\ensmathdix}{msbm10}
\newfont{\ensmathneuf}{msbm10 scaled 833}
\newfont{\ensmathhuit}{msbm10 scaled 694}
\newfam\ensmathfam                        
\textfont\ensmathfam=\ensmathonze        
\scriptfont\ensmathfam=\ensmathdix       
\scriptscriptfont\ensmathfam=\ensmathhuit

\def\be{\begin{equation}}
\def\ee{\end{equation}}
\def\bea{\begin{eqnarray}}
\def\eea{\end{eqnarray}}

\def\d{{\rm d}}

\begin{document}

\selectlanguage{english}

\title{\bf {The Local Time Distribution of a Particle Diffusing on a Graph}}

\author{Alain Comtet$^{1,2}$, Jean Desbois$^{1}$ and  Satya N. Majumdar$^{3}$}

\maketitle	

{\small
\noindent$^{1}$
Laboratoire de Physique Th\'eorique et Mod\`eles Statistiques,
Universit\'e Paris-Sud, B\^at. 100, F-91405 Orsay Cedex, France.

\noindent$^{2}$
Institut Henri Poincar\'e, 11 rue Pierre et Marie Curie, 75005 Paris, France.

\noindent$^{3}$
Laboratoire de Physique Quantique, Universit\'e Paul Sabatier, 31062
Toulouse C\'edex, France.
}
\vskip1cm
\begin{abstract}
\vskip0.5cm
We study the local time distribution of a Brownian particle diffusing
along the links on a graph. In particular, we derive an analytic expression of
its Laplace transform in terms of the Green's function on the graph. 
We show that the asymptotic behavior of this distribution has non-Gaussian
tails characterized by a nontrivial large deviation function.
  
\end{abstract}

\vskip1cm
Graphs are ubiquitous and fascinating objects \cite{Har}. In equilibrium statistical physics,
the study of various model systems on graphs have provided deeper
insights into how thermodynamic properties depend on the geometry
of the underlying graph or network. For example, the study
of a spin system on a complete graph provides an understanding
of the thermodynamical properties at a mean field level. These 
studies on graphs are often nontrivial, the famous example being the
Sherrington-Kirkpatrick model of Ising spin glasses defined
on a complete graph \cite{SK}. More recently, models of nonequilibrium
statistical physics such as the abelian sandpile model have also
been studied on graphs providing deep insights to the dynamics \cite{DD}.
The simplest dynamical example that has evoked enormous interest in both the
physics and the mathematics community is the study of random walks
on graphs, where a particle hops from one vertex to another
provided they are connected by an edge. This has led to the
study of the spectral properties of the discrete Laplacian operator
on both regular \cite{spectral} and random graphs. All these systems
mentioned above share one common property: the links between vertices
on the graph do not play an {\em active} role in the actual physical process, their
only purpose is to provide just a connection between two vertices. For example,
the Euclidean length of a link is completely irrelevant and the physical
properties only depend on parameters such as the coordination number of
vertices etc. Due to this {\em passive} nature of the links, we refer
to such networks as `passive-link' networks.

On the other hand, there exist `active-link' networks where the links
participate directly in the actual physical process. Examples include
the networks for supplying household utilities such as electricity,
water, telephone etc, the network of pipelines carrying oil and natural gases,
the network of blood vessels in a living organism and many others. 
Quantum transport through mesoscopic networks provides yet another
example of such `active-link' network which has been
studied experimentally and theoretically \cite{qt,TM}.
Unlike their passive counterparts, the properties of
physical observables on `active-link' networks do depend on the
Euclidean lengths of the links in addition to other
parameters of the graph. Given the fact that the study of 
random walks on `passive-link' networks
has evoked so much interest and found numerous applications, it is
natural to extend these studies on `active-link' networks. In contrast
to the `passive-link' networks, the random walker does not hop between
vertices of an `active-link' network but actually undergoes continuous
time Brownian motion along the links. A biological example of such
a diffusion process on an `active-link' network is the spread of an 
infectious virus or bacteria along the blood vessels in a living organism.  

In this Letter we study analytically the properties of a 
particular physical observable associated with the diffusion process 
on an `active-link' graph, namely the probability distribution
of the local time spent by the Brownian particle at a given
point on the graph. This quantity characterizes the amount of time $T$ 
spent by the particle between $0$ and $t$ in the vicinity of a point.
The corresponding quantity for a discrete random walk on a `passive-link'
network is the widely studied `number of returns' to a given vertex.
In the biological context mentioned above, the local time $T$ 
denotes the time spent by the diffusing virus in the vicinity
of a point (e.g, near the brain or the lungs) within its
own lifetime $t$ and hence is a measure of the damage that
the virus can cause at a particular place in the network
of blood vessels. Thus the study of the statistical properties 
of the local time in this context is important for medical purposes. 

For a fixed total time $t$, the local time $T$ at any given point on the graph
is clearly a random variable taking a different value for each history
of the diffusion process. In this Letter we study analytically the
probability distribution $P(T,t)$ of the local time and show that 
for any generic graph where the total lengths of the links is finite,
the distribution has the generic asymptotic behavior, $P(T,t)\sim \exp\left[-t\Phi (T/t)\right]$
in the scaling limit $T\to \infty$, $t\to \infty$ but keeping the ratio $T/t$ fixed.
The function $\Phi(x)$ is a large deviation function that characterize the
non-Gaussian tails of this distribution. We provide a general formula
for this large deviation function $\Phi(x)$ and calculate it
explicitly in few specific examples.

Let us begin by
establishing our notation.

\vskip.2cm
  Consider a graph $ \cal G $ consisting of a set of $V$ vertices,
  labelled from $0$ to $V-1$ and linked by $B$ links of finite lengths.
   The coordination   
of vertex $\alpha$ is denoted by  $m_{\alpha }$, therefore
 $\sum_{\alpha =0 }^{V-1} m_{\alpha }= 2 B$.

\vskip.2cm
\noindent
 Each link [$\alpha\beta$]  of length
     $l_{\alpha\beta }$ is identified with an
    interval $[0, l_{\alpha\beta }]$ of $R$. We denote by $x_{\alpha\beta }$
    the coordinate on the link [$\alpha\beta$] starting from vertex $\alpha$ .
 Unless otherwise stated  the
 total length of the graph  $l=\sum_{[\alpha\beta ]}l_{\alpha\beta }$ 
 is assumed to be finite.

\vskip.2cm

 Now, let us consider a Brownian particle starting at $t=0$
 from some point $O$. Without loss of generality we may
assume that this point is a
 vertex ($\alpha=0$) of the graph. As time evolves the particle will
 diffuse along the links and explore the whole graph. If the diffusion
 is recurrent (which is, in particular,  the case for a finite graph), it will
 revisit infinitely often the initial vertex. The aim of this letter  is to compute the
 probability distribution of the total time $T$ spent at $O$ between $0$ and $t$.
 If $x(t)$ denotes the position of the
 particle at time $t$, the local time $T$
 spent by this particle in an infinitesimal neighbourhood of $O$ may be
 defined as
 
 \be\label{d1}
 T \equiv \int_0^t \delta  \left( x(t') \right) \; \d t' ,
\ee
 where $x=0$ is the location of O.
\vskip.2cm
Let $P(T,t)$ be the probability distribution of $T$. This distribution
was studied recently in the one dimensional Sinai type models
and it was shown that the characteristic function of the local
time distribution can be expressed in terms of the evolution of
a quantum Hamiltonian by using a path integral or equivalently
a backward Fokker-Planck approach \cite{maco}. This method
can be easily extended to the present case where one dimensional
lines connect up to form the `active-link' network. We find  
\be
   \ E( e^{-pT})  \equiv
  \int_0^{\infty } \d T \;  P(T,t) \; e^{-pT} =
 \int_{\mbox{Graph } } \d x \;
  \left\langle x \left\vert e^{-tH} \right\vert 0 \right\rangle \label{LT1},
\ee
where $E$ denotes the expectation with respect to all Brownian paths over
time $t$ each starting from $O$ and $H$ is a quantum Hamiltonian defined below.
A further Laplace transform in $t$ gives
\be \label{LT2}
 {\cal L} = \int_0^{\infty } \d t \;  e^{-\gamma t} \; 
 \ E( e^{-pT}) =  \int_{\mbox{Graph } } \d x \; 
  \left\langle x \left\vert \frac{1}{H+ \gamma } \right\vert 0 \right\rangle
 \equiv  \int_{\mbox{Graph } } \d x \; G(x),  
\ee
\noindent
where $G(x)$ is the resolvent.
On each link  $[\alpha\beta ]$,  the  Hamiltonian $H$ acts as the one
dimensional Laplace operator
 $-\frac{1}{2}\Delta $
  ($\equiv -\frac{1}{2}\frac{\d ^2}{\d x_{\alpha\beta } ^2 } $). Moreover,
   the behavior of the resolvent $G(x)$ has to be specified in the
   neighbourhood of all the vertices. 

\vskip.2cm
\noindent
 Consider indeed
 some vertex $\alpha$ and its nearest neighbours $\beta_i$,
 $i=1,2,\ldots ,m_{\alpha }$. Suppose that the Brownian
 particle reaches $\alpha$. It will leave this vertex along the link
 $[\alpha\beta_i]$  with
 some
  probability $p_{\alpha\beta_i }$ that we may choose arbitrarily.
 For  the sake of simplicity, we make the homogeneity assumption
  $p_{\alpha\beta_i } =1/ m_{\alpha }$. With this assumption, the resolvent
  $G(x)$ is shown to be continuous at all the vertices. We will
  denote by $G_{\alpha }$ 
   its value on vertex $\alpha $.
 
\vskip.2cm
\noindent
 If $\alpha\ne 0$, probability conservation implies

\be\label{c1}
\sum_{i=1}^{m_{\alpha } } \frac{\d G}{\d x_{\alpha\beta_i } }
 \bigg\vert_{x_{\alpha\beta_i }=0} \; = \; 0.
\ee

\vskip.1cm
\noindent
In addition, in an infinitesimal neighbourhood of the starting vertex $O$,
 $G$ must satisfy
 
\be\label{c2} 
\left( -\frac{1}{2} \; \Delta +\gamma + p \; \delta   \right) G =  \delta.   
\ee 
This equation is a generalization of the one dimensional case studied in Ref. \cite{maco}.
\vskip.1cm
\noindent
Denoting by $\mu_i$  one of the
nearest neighbours of $O$,
spatial integration in this neighbourhood gives

\be\label{c3} 
-\frac{1}{2} \; \sum_{i=1}^{m_0 } \frac{\d G}{\d x_{0\mu_i } }
 \bigg\vert_{x_{0\mu_i }=0} \; + \; p \; G_0 \; = \; 1 .
\ee
\noindent
Let us now show that all the derivatives of $G$ appearing in the above
equations can be expressed in terms of the $G_{\alpha }$'s. On the link
  $[\alpha\beta ]$,  $G(x_{\alpha\beta })$  satisfies 

\be\label{der1}
 \left( -\frac{1}{2} \; \frac{\d ^2}{\d x_{\alpha\beta } ^2 } \; + \; \gamma 
 \right) G(x_{\alpha\beta }) =0,
\ee
whose solution is
\be\label{der2}
 G(x_{\alpha\beta })= G_{\alpha } \; 
\frac{ \sinh \sqrt{2\gamma }(l_{\alpha\beta } - x_{\alpha\beta })}
     { \sinh \sqrt{2\gamma } l_{\alpha\beta } } +
 G_{\beta } \; 
\frac{ \sinh \sqrt{2\gamma } x_{\alpha\beta } }
     { \sinh \sqrt{2\gamma } l_{\alpha\beta } }. 
\ee

\vskip.1cm
\noindent
It follows that
\be\label{der3}
\frac{\d G}{\d x_{\alpha\beta } }
 \bigg\vert_{x_{\alpha\beta }=0} = -  c_{\beta\alpha } \; G_{\alpha } +
 s_{\alpha\beta } \; G_{\beta } ,
\ee
where
\bea
c_{\alpha\beta } &=& \sqrt{2\gamma } \; 
 \coth \sqrt{2\gamma } l_{\alpha\beta } = c_{\beta\alpha } \label{der4} \\
  s_{\alpha\beta } &=& \frac{  \sqrt{2\gamma }}{
  \sinh \sqrt{2\gamma } l_{\alpha\beta }} =  s_{\beta\alpha }.
   \label{der5}
\eea

\noindent
The relation (\ref{der3}) allows us to write Eqs. (\ref{c1},\ref{c3}) in a matrix form

\be\label{mat1}
M(p) \; G=L,
\ee
where $M(p)$ is a symmetric $(V \times V)$ matrix with elements

\bea
 M_{00}&=&\sum_{i=1}^{m_0}c_{0\mu_i} +2p   \label{m1} \\
 M_{\alpha\alpha }&=&\sum_{i=1}^{m_{\alpha }}c_{\alpha\beta_i} \quad
 \mbox{if} \quad \alpha\ne 0
  \label{m2} \\
M_{\lambda\mu } &=& - s_{\lambda\mu } \quad \mbox{if} \quad
 [\lambda\mu ] \quad  \mbox{is a bond} \label{m3} \\
  &=& 0 \qquad \mbox{otherwise}. \label{m4}
\eea

\noindent
$G$ and $L$ are two $(V\times 1) $ matrices whose elements are, respectively,
 $G_{\alpha }$ and $L_{\alpha }= 2 \delta_{\alpha 0}$. Inverting Eq. (\ref{mat1})
we get 
\be\label{r1}
G_{\alpha }=2 \left( M(p)^{-1}  \right)_{\alpha 0}.
\ee
Substituting the results from Eqs. (\ref{der2}) and (\ref{r1}) in the original expression
in Eq. (\ref{LT2}) and carrying out a few elementary manipulations on determinants, we get
a closed form expression for the (double) Laplace transform,
\be\label{r2}
 {\cal L} =\frac{1}{\gamma } \; \frac{\det M(0)}{\det M(p)}. 
\ee
To proceed further, 
 let us denote by $\widehat{G_{\alpha }}$  the value of $G_{\alpha }$
 for $p=0$. Due to the special form of the matrix $M(p)$, we may
 write
\be\label{r3} 
\det M(p) = (1+p \; \widehat{G_0} ) \; \det M(0)
\ee
and recast (\ref{r2}) in the form
\be\label{r4}
\int_0^{\infty }  \d T \; e^{-p T} \int_0^{\infty }  \d t \; 
e^{-\gamma t} \; P(T,t) \; = \; \frac{1}{\gamma } \;
 \frac{1}{1+p \; \widehat{G_0} }.
\ee

\vskip.2cm
\noindent
Since $ \widehat{G_0}  $ is independent of $p$, we may invert the Laplace
transform with respect to $p$ and get the relationship
 
\be\label{r5} 
 \int_0^{\infty }  \d t \; e^{-\gamma t} \; P(T,t) =
\frac{\lambda (\gamma )}{ \gamma } e^{-T \lambda (\gamma ) }
\ee
with
\be\label{r6} 
\lambda (\gamma ) = \frac{1}{ \widehat{G_0}  }.
\ee

\vskip.2cm
\noindent
Eq. (\ref{r5}) is the central result of this Letter. It provides a
generalization to graphs of the L\'evy formula for one dimensional
diffusion processes \cite{L}. It is worthwhile to notice that 
 Eqs. (\ref{r2},\ref{r5},\ref{r6})
 are valid for all kind of graphs irrespective of whether they are finite or not.
 The simplicity of this result calls for an alternative derivation based on
 paths integrals. The starting point is a representation of the resolvant as
 a two point function of a scalar field theory where the field $\phi(x)$ is
 a scalar field defined on the links of the graph and satisfying suitable
 continuity properties (\cite {AC}). 
\be\label{r7}
\left\langle x \left\vert \frac{1}{H+ \gamma } \right\vert 0
\right\rangle= Z(p)
\ee 
where
\be\label{t8}
 Z(p) = \frac{\int{\cal D }\phi\phi(x)\phi(0)
 \exp-{ S (\phi)}}{\int{\cal D }\phi\exp-{ S (\phi)}}
\ee
and the action is 
\be\label{t9}
S(\phi)= \frac{p}{2}\phi(0)^2+\frac{1}{2}\int\d x (\frac{1}{2}(\nabla\phi)^2
+\gamma\phi^2)
\ee
One first expands the exponential term in $p$ both in the numerator and
denominator and computes all contractions
using Wick's theorem. Then one can resum the series and obtain after some
algebra

\be\label{t10}
\left\langle x \left\vert \frac{1}{H+ \gamma } \right\vert 0
\right\rangle= \frac{ \widehat{G_x}}{1+p \; \widehat{G_0} }.
 \ee
Substituting this result in Eq. (\ref{LT2}) and integrating over $x$, 
we get back the same result as in Eq. (\ref{r6}).
 
\vskip.6cm
The computation of the local time distribution is therefore reduced to the
calculation of the Green's function on the graph. In some cases it is quite
useful to express $\lambda (\gamma)$ as a ratio of two spectral
determinants of the form
\be\label{t11}
\lambda(\gamma)=\frac{1}{2} \frac{S_{N}(\gamma)}{S_{D}(\gamma)},
 \ee
where $S(\gamma)=\det(-\triangle+\gamma)$ is the spectral determinant
of the graph. The numerator is computed with Neumann boundary
conditions at all vertices and the denominator is computed with Dirichlet
boundary conditions at O and Neumann at all other vertices.
Note that such ratios of spectral determinants which arise
in the context of scattering theory on graphs \cite{AC,D1,D2} have 
recently appeared in the
mathematical litterature in relation with  Dirichlet forms \cite{S} or
Sturm-Liouville problems \cite{V}. 

In order to extract the behavior of the distribution $P(T,t)$ from
the general formula in Eq. (\ref{r5}), we need to compute the function $\lambda(\gamma)$
explicitly for a given graph and then invert the Laplace transform. In general,
it is not easy to invert this Laplace transform. However, in the asymptotic limit
when both $t$ and $T$ large with their ratio fixed, it is possible to
make progress. In this scaling limit, one expects, from generic considerations,
the following scaling behavior \cite{maco,MB},
\be\label{sl1} 
 P(T,t) \sim e^{-t \Phi (T/t)},
\ee
where $\Phi(y)$ is a large deviation function. Substituting this expected scaling form
on the left hand side of Eq. (\ref{r5}) and carrying out a steepest descent
calculation valid for large $t$ and $T$, one obtains the following Legendre
transform,
\be\label{sl2} 
\Phi (y) = \max_{\gamma } \left[  -\gamma +y \; \lambda (\gamma )   \right].
\ee

Thus the large deviation function can be computed from Eq. (\ref{sl2})
provided one knows the function $\lambda(\gamma)$ explicitly for a given
graph. Examples where one can calculate explicitly the function
$\lambda(\gamma)$  are given below.
Interestingly, however, some general
features of the large deviation function
 $\Phi(y)$, such as its behaviour
near the tails as well as near its minimum can be worked out for
a generic graph even without
the detailed knowledge of the function $\lambda(\gamma)$, thus displaying
certain universal features.
 
\vskip.4cm
First, let us consider finite graphs. One can show that $\Phi (y) $ decreases
monotonically with increasing $y$ in the range $0<y<1/l$, achieves a
minimum at $y=1/l$ and then increases monotonically for $y>1/l$ (recall that $l$ 
is the total length of the graph). We consider the following three regions
where the behavior of $\Phi(y)$ is quite generic. 

\vskip.2cm 
\noindent
i) Let us first consider the limit $\gamma \to \infty$.
Using Eqs. (\ref{m1}-\ref{m4}) and (\ref{r3}), we get 
\be\label{sl4} 
 \lambda (\gamma ) \sim m_0 \; \sqrt{\frac{\gamma }{2}},
\ee
where $m_0 $ is the coordination of the vertex $O$. Using this expression in Eq. (\ref{sl2})
and maximizing, we get
\be\label{sl5}
 \Phi (y) \approx \frac{m_0^2 \;  y^2}{8}  \quad
 \mbox{when}  \quad y\to \infty.
\ee

\vskip.1cm 
\noindent
The limit $ y\to \infty $ means that we select infinitely long Brownian trajectories
 (recall that $ t\to \infty $) for which the particle stays during almost all the
  time close to $O$. Since the particle does not  explore the whole
  graph, the local time distribution will    
   mainly depend on the characteristics of the graph in the vicinity of
   $O$, i.e. on the coordination $m_0$.  

\vskip.2cm 
\noindent
ii) By considering the first negative singularity, $\gamma_0 $, of
  $ \lambda (\gamma ) $  in Eq. (\ref{r6}) we can show that, for
   $\gamma \approx \gamma_0 $,
   $ \lambda (\gamma ) \sim a/(\gamma_0 -  \gamma ) $ with $a>0$. From 
 Eq. (\ref{sl2}) we deduce that
\be\label{sl3}  
\Phi (y) \approx \vert \gamma_0 \vert - 2 \sqrt{ay} \quad
 \mbox{when} \quad y\to 0^+
\ee
 where $\gamma_0$  is the lowest eigenvalue of the Laplacian on the graph
 with Dirichlet boundary conditions in O.
\vskip.2cm 
\noindent
iii) Finally, it turns out that $\gamma \to 0$ limit corresponds, via Eq. (\ref{sl2}), to
$y\to 1/l$. In this limit
\be\label{sl6} 
\Phi (y) \approx \frac{1}{C} \left(y- \frac{1}{l} \right)^2
 \quad
 \mbox{when}  \quad y\approx \frac{1}{l}. 
\ee

\noindent
The positive constant $C$ is obtained by a small $\gamma $ expansion
in Eq. (\ref{mat1}) with the result: 
\be\label{sl7} 
C= \frac{2}{ l^3} \; \left\vert \frac{\det Q}{\det N} \right\vert,
\ee
where the matrix $N$ is obtained by removing  the $1$st line and  $1$st column
 in the matrix $M(0)$  and  taking the limit $\gamma \to 0$.
The matrix $Q$ is obtained by  taking the limit $\gamma \to 0$ in the matrix
 $M(0)$
 and replacing the  $1$st line and  $1$st column by the elements
 $Q_{00}= -(1/3)\sum_{[\alpha\beta ]} l_{\alpha\beta }^3$ and
 $Q_{0\alpha}= Q_{\alpha 0 }= \sum_{i=1}^{m_{\alpha }}l_{\alpha\beta_i }$ for
 $\alpha\ne 0$.
 
\vskip.3cm
\noindent
Note that the value $1/l$ corresponds to the mean local time $\langle T\rangle/t$  and the constant
$C$ is proportional to the variance of the local time. This result, which
also follows from a central limit theorem, indicates a Gaussian distribution
for the local time near its mean value,
 $P(T,t)\sim \exp\left[-(T-t/l)^2/{2\sigma^2}\right]$,  
where the variance, $\sigma^2 = Ct/2$ for large $t$. This immediately
gives a large deviation function $\Phi(y)$ as in Eq.(\ref{sl6}). Note, however,
that this central limit theorem holds only near $y\approx 1/l$ but fails
in regions away from $y=1/l$. In particular, near the tails the distribution
becomes non-Gaussian as reflected in the form of $\Phi(y)$ near
$y\to 0$ and $y\to \infty$ above. For one dimensional diffusion processes,
these characteristics of large deviation functions were identified not just
for the local time but for other functionals of Brownian motion as well 
 \cite{maco,MB}.

\begin{figure}[!ht]
\begin{center}
\includegraphics[scale=0.9]{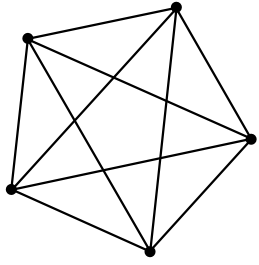}
\end{center}
\caption{The complete graph $K_5$.}
\end{figure}
\vskip.4cm 
\noindent 
   As an example, consider a complete graph $K_{n}$
with $n$ vertices where there is a link of length $d$ between any pair of vertices (see
fig.1). 
In this case, putting $p=0$ in Eq. (\ref{r1}), evaluating the inverse
of the $M$ matrix explicitly and using this result in Eq. (\ref{r6}), 
one obtains (\cite {AC}) 
 \be\label{t12}
 \lambda(\gamma)=\frac{1}{2}
 \frac{\sqrt{2\gamma}}{\cos(\phi)} \frac{\left[\cosh(d\sqrt{2\gamma})+\cos
 (\phi)\right]}{\left[(\cosh(d\sqrt{2\gamma})+\cos(\phi)-1)\right]} 
\tanh\left[\frac{d}{2}\sqrt{2\gamma}\right]
 \ee
where $\cos(\phi)=\frac{1}{n-1}$. In particular, a small $\gamma $ expansion
 in (\ref{sl2}) leads for the constant 
 $C$ to the value $C= 16(n+1/n-11/6)/(n(n-1))^2$,
 in agreement with Eq. (\ref{sl7}).

\vskip.4cm 

Another interesting example, which was recently used in the context of
quantum chaos \cite {BB}, \cite {KS}  is the case of a star graph consisting
of $n$ links of length $d$ attached to the vertex O (fig 2).
\begin{figure}[!ht]
\begin{center}
\includegraphics[scale=0.45]{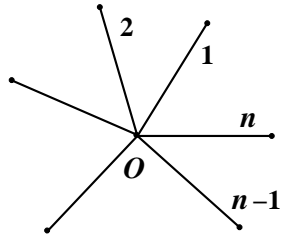}
\end{center}
\caption{The hydra with $n$ legs.}
\end{figure}

\noindent
In this case one obtains
\be\label{u1}
\lambda(\gamma)=n\sqrt{\frac{\gamma}{2}}\tanh(d\sqrt {2\gamma} )
\ee
One can also compute explicitly the tails of the large
deviation function and thus check that they are in agreement with the general
results Eqs. (\ref{sl5}-\ref{sl7}). 
Clearly, in the latter cases ii) and iii), the whole graph is explored 
 and the distribution depends, in a rather nontrivial way, on the
 details of the graph.

\vskip.6cm 

The previous formulae can be easily generalised to infinite graphs.
The example of a star graph with infinite legs is
interesting since it was already used in a probabilistic setting  to derive 
multidimensional
extensions  of the arc sine law (\cite {B}). Using 
Eq. (\ref {u1}) one can derive  the exact  probability distribution

\be\label{u2} 
P(T,t)= \frac{n}{\sqrt{2\pi t} }\exp\left[- \frac{n^2 T^2 }{8 t}\right]
\ee
The fact that the distribution is purely Gaussian depends obviously heavily
on the symmetries of the graph. However the behaviour for small and large $T$
is  universal. 
 For an arbitrary graph with $n$ infinite legs attached in any vertex 
one can show
that
 $ \Phi (y) $ is a monotonically increasing function for $y>0$ and also
  that

\vskip.2cm 
\noindent
i)
\be\label{sl50}
 \Phi (y) \approx \frac{m_0^2 \; y^2}{8}  \quad
 \mbox{when}  \quad y\to \infty
\ee
(same result as before and for the same reasons) 

\vskip.2cm 
\noindent
ii)
\be\label{sl51}
 \Phi (y) \approx \frac{n^2 \; y^2}{8}  \quad
 \mbox{when}  \quad y\to 0
\ee
(the particle tries to explore the whole graph; so, it essentially
 ``sees'' the infinite legs).

\end{document}